\documentclass{raa}            % referee version: for submission

\usepackage{graphicx,times}             %for PS/EPS graphics inclusion, new
\usepackage{natbib}
\usepackage{amssymb,amsmath}

\bibpunct{(}{)}{;}{a}{}{,}

\usepackage[pagebackref=true]{hyperref}

\begin{document}

  \title{The Mini-SiTian Array: White Paper
}
%   \subtitle{I. Place Your Subtitle Here}

   \volnopage{Vol.0 (20xx) No.0, 000--000}      %%preserved for Editor. DOn't remove!
   \setcounter{page}{1}          %%starting page, preserved for Editor. DOn't remove!

   \author{Henggeng Han %(韩恒赓) %% Put your Chinese name in "( )" if you like. Note to open line 11 "\usepackage[UTF8]{ctex}"
      \inst{1}
   \and Yang Huang
      \inst{2, 1}
   \and {Beichuan Wang}
      \inst{1, 2}
   \and Yongkang Sun
      \inst{1, 2}
   \and Cunshi Wang
      \inst{2, 1}
   \and{Zhirui Li}
      \inst{1, 2}
   \and{Junjie Jin}
      \inst{1}
   \and{Ningchen Sun}
      \inst{2}
   \and{Kai Xiao}
      \inst{2}
   \and{Min He}
      \inst{1}
   \and{Hongrui Gu}
      \inst{1, 2}
   \and{Zexi Niu}
      \inst{2}
   \and{Hong Wu}
      \inst{1}
   \and{Jifeng Liu}
      \inst{1, 2, 3, 4}
   }

   \institute{National Astronomical Observatories, Chinese Academy of Sciences,
             Beijing 100012, China; jfliu@bao.ac.cn \\
        \and
             School of Astronomy and Space Science, University of Chinese Academy of Sciences, Beijing 100049, People's Republic of China; huangyang@ucas.ac.cn\\
        \and
             Institue for Frontiers in Astronomy and Astrophysics, Beijing Normal University, Beijing, 102206, People's Republic of China\\
        \and
            New Cornerstone Science Laboratory, National Astronomical Observatories, Chinese Academy of Sciences, Beijing, 100012, People's Republic of China
\vs\no
   {\small Received 20xx month day; accepted 20xx month day}}

\abstract{This paper outlines the scientific goals and observational strategies of the Mini-SiTian array. Mounted at Xinglong Observatory, the Mini-SiTian array consists of three 30 cm telescopes and has been in operation since 2022. The large field of view, combined with the capability for multi-band photometric observations, enables the Mini-SiTian array to perform rapid follow-up observations to identify optical counterparts of gravitational waves, capture the early light curves of tidal disruption events and supernovae, and monitor stellar flares, Be star outbursts, and cataclysmic variable stars, although its limiting magnitude is not very deep. By collaborating with the Xinglong 2.16-m telescope and leveraging a real-time image processing pipeline, simultaneous photometric and spectroscopic observations could be performed to reveal their underlying physical mechanisms. The observational and research experience provide critical guidance for the implementation of the full-scale SiTian project in the future.
}

   \authorrunning{Henggeng Han}            %author_head in even pages
   \titlerunning{Sciences of Mini-SiTian}  % title_head in odd pages

   \maketitle
%% The author head (on even pages) and the title head (on odd pages) will be
%% automatically extracted from \author{} and \title{}. Whenever the title is too long,
%% you will be asked to supply a shorter one by inserting either \authorrunning{} or
%% \titlerunning{} before \maketitle. Anyway, you can specify your own heads.
%%
%%
%% Note: In the following text body of your manuscript, please note several differences from
%%       other major journals:
%% (1) \subsection{Please Capitalize the First Letter of Each Notional Word in Subsection Title}
%% (2) Please Capitalize the First Letter of Each Notional Word in all tables' captions

%
%________________________________________________ sections below
%
\section{Introduction}           %% first-level sections will be auto-capitalized
\label{sect:intro}

In our universe, there are many types of variable stars including rotating variables, eruptive variables, pulsating variables, which are important tracers to understand the universe. For example, classical Cepheids and Type Ia supernovae could be served as the standard candles \citep[e.g.][]{1912HarCi.173....1L, 1998AJ....116.1009R}, flaring stars are keys to shed light on stellar dynamo theories \citep{2014ApJ...797..121H, 2016ApJ...829...23D} and multi-messenger observations reveal that the gravitational wave event GW1770817 is a result of neutron star$-$ neutron star merger \citep{2017PhRvL.119p1101A}.

Long-term observations and immediate follow-up observations are required to reveal the nature of the variability \citep{2000PASP..112.1281P}. As a result, time-domain astronomy, which seeks to map the universe’s dynamic changes over timescales ranging from hours to years, is now a rapidly growing field. Thanks to the development of modern techniques, large ground-based and space-borne telescopes gradually came into being, which opens the new era of time-domain astronomy. Up to now many remarkable time-domain sky surveys have been carried out, including the Catalina Real-Time Transient Survey with a field of view (FOV) of 19.4 $\rm{deg^{2}}$ and a limiting magnitude of $V\sim19.5$ \citep{2009ApJ...696..870D}, the All-Sky Automated Survey for Supernovae (ASAS-SN) with very large sky coverage and a limiting magnitude of $V\sim 17$ \citep{2014ApJ...788...48S, 2017PASP..129j4502K} and the Zwicky Transient Facility (ZTF) time-domain survey with FOV of 47 $\rm{deg^{2}}$ and a limiting magnitude of $\sim 21.1$ \citep{2019PASP..131a8002B}. 

Meanwhile, there are some forthcoming and onging time-domain sky surveys including the Wide field survey telescope \citep[WFST;][]{2023SCPMA..6609512W}, the Large Synoptic Survey Telescope \citep[LSST;][]{2019ApJ...873..111I}. The 1.6m Multi-channel Photometric Survey Telescope of Yunnan University with FOV of 3.14 square degree \citep[Mephisto;][]{2020SPIE11445E..7MY}. After being fully operational, these surveys would provide us with tens of thousands light curves, which are essential for searching for the variables and investigating their nature.

However, these surveys are limited to their relatively long cadence. In this regard, they would miss the early stages of many transients events. Meanwhile, it is difficult for them to trigger immediate follow-up spectrosocpic observations. Acknowledging these shortcomings, \cite{2021AnABC..93..628L} are now carrying out an innovative time-domain sky survey project, named SiTian (a Chinese character which means monitoring of the sky) project.
SiTian project will consist of $\sim$ 20 nodes with each node containing three 1-m telescopes. In other words, SiTian project will contain 60 1-m telescopes. For each node, the three telescopes are equipped with \emph{g}, \emph{r}, or \emph{i} band filters, respectively, which could enable multi-band observation simultaneously. After being fully operational, the total coverage corresponding to one exposure will be 600 $\rm{deg^{2}}$ for a node with a limiting magnitude of 21.0 mag at a cadence of 30 minutes, which means a total coverage of 12,000 $\rm{deg^{2}}$ for 20 nodes. Meanwhile, SiTian project will be equipped with at least one 4-m telescope, aiming in carrying out follow-up spectroscopic observations of interesting targets. For more details of SiTian project, please refer to \cite{2021AnABC..93..628L}.

%\sunnc{Mini-SiTian}

Inspired by the concept of the SiTian project, we designed the pilot project named Mini-SiTian project, whose main purpose is to refine the technical details of the SiTian project and to conduct preliminary observations for the scientific program of the SiTian project. Acting as the pioneer of the SiTian project, the Mini-SiTian array has already been installed at the Xinglong Observatory. The array consists of three 30-cm telescopes with each capable of functioning in either the \emph{g}, \emph{r}, or \emph{i} band. Each telescope has a FOV of $\sim$ 3 $\rm{deg^{2}}$. For technique details of Mini-SiTian we refer to the paper of the series by Han et al. (2025). The Mini-SiTian array has being operated since September of 2022 and the data have been used to test the performance of the telescopes. For more details we refer to the paper of the series by He et al. (2025).

Both the immediate follow-up observations of transients and regular sky survey will be performed by the SiTian project. As the pathfinder of SiTian project, Mini-SiTian also follows the same observation strategies. Catching the early stages of transients as well as regular sky survey will be performed. Such strategy would be suitable for searching for electromagnetic (EM) counter part of gravitational wave (GW), catching the early light curves of supernovae and tidal disruption events and identifying variables including flaring stars, Be outbursts and cataclysmic variable stars. Meanwhile, the artificial intelligence could act as a powerful assistant, which could improve the observing strategies of Mini-SiTian. In this paper, we will describe the scientific goals of Mini-SiTian in detail. 

\section{Scientific Goals}
\subsection{Electromagnetic counterpart of gravitational wave}

The successful detection of gravitational wave GW170817 and its electromagnetic counterpart has opened the new era of multi-messenger astronomy of gravitational wave \citep{2017PhRvL.119p1101A}. The electromagnetic counterpart transient of GW170817 is attributed to a kilonova model, which is consistent with r-process of nucleosynthesis \citep{2017Natur.551...75S}. Kilonova is the brightest electromagnetic transient of gravitational wave \citep{2013ApJ...774...25K}. As a result, catching the early stage light curves of kilonova is of great importance while studying the equation of state and mass distribution of neutron stars. 

One of the major scientific goal of SiTian-project is to apply immediate follow-up observations if gravitational wave events occur. Due to the limitation of FOV of Mini-SiTian array, the observing strategy would be quite different from the one of SiTian project. Once received an alert from the LIGO regarding a gravitational wave event, our scientific group will judge whether such event is worth observing. If so, the ongoing observation will be interrupted and follow-up observations will be carried out immediately. Finding the electromagnetic counterpart of gravitational wave has the highest priority. For more details of the capability of searching for kilonova events based on SiTian project, we refer to the paper of the series by Li et al. (2025).

\subsection{Flaring stars}
\subsubsection{Background}
%\subsection{Flaring stars}

Stellar flares are believed to be due to reconnection of stellar magnetic fields, during which a large amount of energy would be released. This leads to brightening of the star in various bands including radio emission \citep{2002ARA&A..40..217G}, optical flares \citep[e.g.][]{2014ApJ...797..121H, 2017ApJ...849...36Y}, UV flares \citep[e.g.][]{2019ApJ...883...88B} and X-ray flares \citep[e.g.][]{2022ApJ...938...76G}. Typically, for optical flares the energy release could reach $10^{29}$ to $10^{32}$ erg \citep{2017ApJ...849...36Y} or even higher than $10^{34}$ ergs for superflares observed in those most active stars \citep{2012Natur.485..478M}.

The solar-stellar connection has long been a hot topic among stellar activity researches. \cite{2012Natur.485..478M} studied the flare frequency distribution of superflares detected on solar-like stars in the \emph{Kepler} field. They argued that on solar-like stars, i.e, slowly rotating G-type stars with effective temperatures of $\sim$ 5700 K, flares with energy higher than $10^{35}$ erg could only happen once every 5000 years. \cite{2017ApJ...848...41L} suggested that extremely powerful superflares could strongly affect habitability of surrounding planets and superflares may occur on the Sun within $\sim$ 1000 years. 

\begin{figure}
\centering
\includegraphics[width=0.5\textwidth]{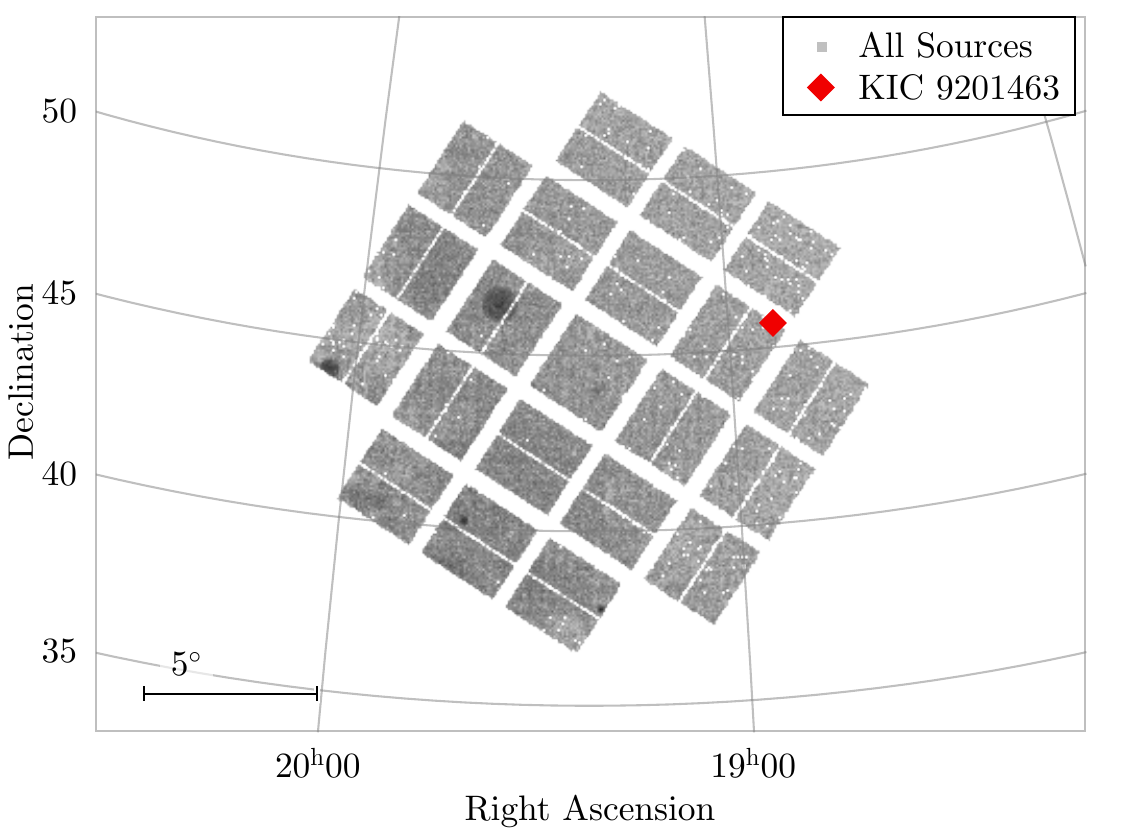}
\caption{Kepler field. KIC 9201463 is marked with a red diamond, which has been observed by the three Mini-SiTian telescopes simultaneously.}
\label{HRD.fig}
\end{figure}

Statistical study of stellar flares could shed light on stellar structure, stellar dynamo and stellar habitability. One of the key projects of Mini-SiTian is to observe flaring stars in the \emph{Kepler} and TESS field, especially late-type stars with strong magnetic fields, on which frequent and huge flares could be detected. Fortunately, about 3400 flaring stars have been identified in the \emph{Kepler} field \citep{2019ApJS..241...29Y}.  Meanwhile, during the TESS first and second years' observations, many bright stars with superflares have been discovered \citep[e.g.][]{2021ApJS..253...35T}. These are ideal sample to carry out continuous multi-band follow-up observations. 

\subsubsection{Observation design}
The Mini-SiTian array has initiated a long-term monitoring project of the \emph{Kepler} field since 2023. This project in combination with the \emph{Kepler} observations could reveal the long-term brightness variations of \emph{Kepler} targets. Many \emph{Kepler} targets exhibit frequent superflares and spectroscopic observations are essential for revealing the energy release process and the mass ejection process during stellar flares \citep[e.g.][]{2021PASJ...73...44M}. Typically, KIC 9201463 is the one of the most active flaring stars in the Kepler field \citep{2019ApJS..241...29Y}, which exhibited stellar flares almost every day and thus increases the probability to capture stellar flares. Meanwhile, the \emph{Kepler} fields are frequently observed by the LAMOST telescope \citep{2020RAA....20..167F}. As a result, for the Mini-SiTian telescopes we propose a long-term monitoring of flaring stars in both the \emph{Kepler} and TESS fields. In Figure 1, we plot the \emph{Kepler} field. The red diamond represents KIC 9201463, which has been observed by the Mini-SiTian array since 2023. Such observations could be crucial for investigating different behaviours of stellar flares, which would be useful for constraining stellar dynamo theories. 

SiTian brain, the main control system of SiTian project, which integrated scheduling observations, real-time data analysis and immediate communication of transients alerts, will act as the central facility \citep{2021AnABC..93..628L}. When transients occur, SiTian brain will coordinate the regular scan mode and the immediate follow-up observation mode. As a pioneer of SiTian project, Mini-SiTian will also be equipped with transients alert system and real-time data processing pipelines. For more technical details we refer to Gu et al. (2025). Such system could inform observers if transients occur in the field and present the early light curve immediately and trigger multi-band follow-up observations. 

Considering the visibility of \emph{Kepler} field at Xinglong observatory, it is much better to monitor such field as long as possible during September and October of each year. The longer the duration is, the more flares could be caught. A longer duration will allow the analysis of flare frequency distribution and its impact on stellar habitable zone. Typical timescales of flares range from minutes to hours \citep{2014ApJ...797..121H, 2017ApJ...849...36Y}. Thus the exposure time is set to be 2-min. A 2-min exposure time would make it possible to reveal more details of flaring phases, which would be useful for investigating topology of magnetic fields. Despite the \emph{Kepler} field, since the TESS satellite is a all-sky survey, it is possible that Mini-SiTian could carry out a simultaneous observation to the TESS fields. Later in 2024, Mini-SiTian started to monitor the TESS fields in which there are solar-like stars with frequent superflares. The observing strategies would be similar despite that the exposure time needs to be reduced. 

\subsection{Tidal Disruption Events and Supernova}
\subsubsection{Tidal disruption events}
When a star is perturbed occasionally and comes into the tidal sphere of a massive black hole (MBH) hosted in the center of its galaxy, it will be tidally disrupted and partially accreted. This process, referred as a tidal disruption event (TDE), accompanied by the emission of bright flare, which decays on timescales of months to years (\citealt{1988Natur.333..523R}). The predicted event rate of TDEs generally falls within the range of $10^{-4}-10^{-5}\text{gal}^{-1}\text{yr}^{-1}$, depending on the properties of the galaxy and the mass of MBH. Recently, Wang et al. (2024), one of the paper in the series, has estimated the detection rates of TDE using the Mini-SiTian telescopes. 

TDEs originated in late 1970s as a theoretical concept. The first X-ray TDE was serendipitously identified from archival \emph{ROSAT} data in 1990s. Subsequently, a couple more TDEs have been discovered with the launch of XMM-Newton and Chandra satellites. Optical TDEs, however, were not identified until 2010s from the archival SDSS data. Thanks to a variety of wide-field optical surveys, such as the ASAS-SN \citep{2014ApJ...788...48S}, the Asteroid Terrestrial-impact Last Alert System \citep[ATLAS;][]{2018PASP..130f4505T} and the ZTF \citep{2021ApJ...908....4V}, an explosively growing number of TDEs have been found in the past decade \citep{2021ARA&A..59...21G}. Consequently, the combination of time-domain surveys and follow-up observations plays a crucial role in the systematic search for TDEs.

TDEs aroused extensive interests because of its scientific values. First of all, TDEs provided us an unique means to probe the SMBHs in quiescent galaxies, which is otherwise particularly difficult in dwarf and distant galaxies. In addtion, TDEs can even probe the dormant intermediate-mass BHs (IMBHs) and supermassive BHs (SMBHs) binaries \citep{2020ARA&A..58..257G, 2021ApJ...920...12H, 2024arXiv240600923H}. The IMBHs are thought to be the seeds from which SMBHs grow, as they link between stellar masses BHs and SMBHs. 
The observed emission of TDEs depends on the parameters such as the black hole mass and spin, hence they offer a unique way to constrain the masses and spin of dormant SMBHs. Moreover, TDEs also act as an ideal laboratory to study the formation of an accretion disk and jet by monitoring the entire life cycle of BH activity. The observations of the infrared and radio echoes in TDEs provide a novel tool to investigate the environment surrounding BHs \citep{2021ARA&A..59...21G}. 

\subsubsection{Supernovae}
The earliest classifications of supernovae (SNe) were based on their different behaviours in spectroscopic observations, which involves whether their spectra exhibit signatures of Hydrogen. Type II SNe show signature of Hydrogen while Type I SNe do not \citep{1941PASP...53..224M}. Meanwhile, light curves of Type I SNe are homogeneity while Type II SNe have light curves that are heterogeneous, which could provide valuable insights into SNe classification, explosion mechanism, progenitor stars and nucleosynthesis processes. 

One of the unsolved problem regarding SNe is the progenitor problem. For Type Ia SNe, Single-Degenerate Models \citep{1973ApJ...186.1007W}, Double-Degenerate Models \citep[e.g.][]{1984ApJS...54..335I}, Collisional Double-Degenerate Models \citep[e.g.][]{1989ApJ...342..986B}, Double Detonations and Rotating Super-Chandrasekhar-Mass Models \citep{1980ApJ...237..142T} and some other alternative models for example ``core-degenerate'' model \citep{2011MNRAS.417.1466K} have been proposed. Light curves at early stages could shed light on these models and provide strong constraints. As for Type II SNe, the variations of their light curves and spectra could reveal the masses of their progenitors, the mass ejection process and energy source \citep[e.g.][]{2009ApJ...703.2205K}. Fortunately, the large FOV of Mini-SiTian arrays, the real-time data processing pipelines together with the Xinglong 2.16-m telescope could make it possible for us to monitor the early light curves of supernova and their spectra in the time-domain, making it possible to investigate the physical models of SNe.  

\subsubsection{Observation design}

\begin{figure}
\centering
\includegraphics[width=1\textwidth]{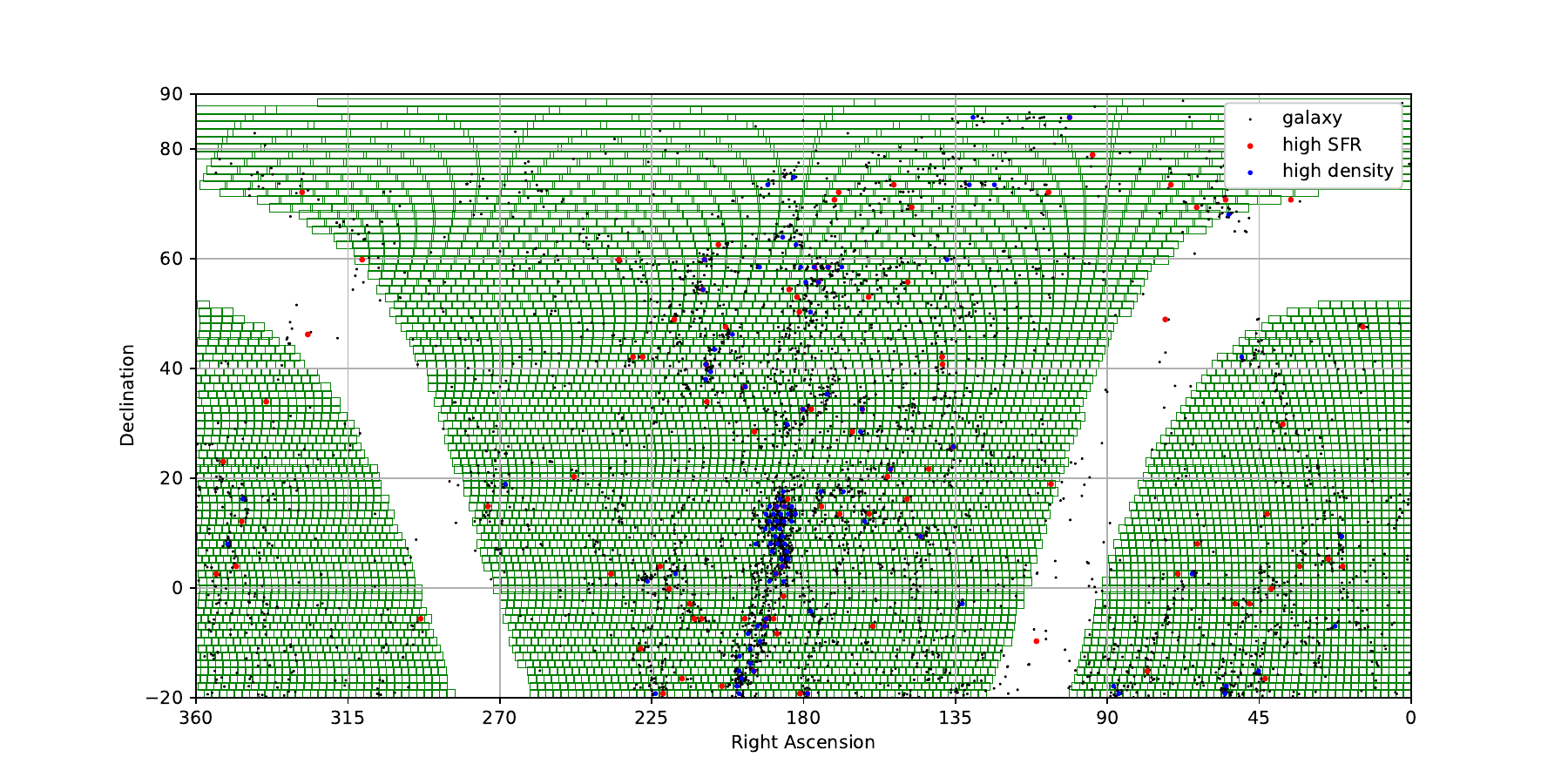}
\caption{Sky region which would be monitored to search for TDEs and SNe. Areas with high galaxy density are marked with blue dots. Areas with high star formation rate are marked with red dots.}
\label{HRD.fig}
\end{figure}

For both TDEs and SNe, we adopt the following survey plan and strategies. (1) The sky region with a declination above $-20$ degrees is segmented into separate areas and a sky region table is established. Each sky region will maintain a certain overlap with surrounding ones, i.e., 1 arcminute. See Figure 2 for more details. They are suitable for searching for TDEs. Star formation rates are higher among areas marked by red dots and they are designed for searching for SNe. (2) For a continuous sky survey, we follow the same strategy as Mini-SiTian regular sky survey with a cadence of 30 minutes, which has a 5 minutes exposure time, circularly observing 1-2 sky regions. For an independent sky region detection, we select sky regions and exposure time based on scientific objectives. (3) Follow-up spectroscopic observations based on the 2.16-m telescope at Xinglong Observatory will be immediately applied once a possible TDE or SN is recognized by the \emph{STRIP} pipeline and our science group.

\subsection{Be stars}
\subsubsection{Background}
Be stars are fast rotating B-type stars that present H$\alpha$ emission lines, which are believed to originate from the viscous decretion disks \citep{2013A&ARv..21...69R}. Many studies have tried to reveal the possible mechanisms of the Be outbursts, including interaction of close-in companion and non-radial pulsations \citep[e.g.][]{2008A&A...484L..13K, 2018A&A...620A.145B, 2020A&A...636A.110D}. The mass outflow could feed the decretion disks and thus leads to enhanced H$\alpha$ emission lines. However, up to now there is no decisive evidence that can distinguish those scenarios. Carrying out multi-band observations of Be stars would be crucial for investigating the nature of decretion disks and mass ejection of Be stars. 

In addition, mass distribution of compact objects is an important piece of binary evolution theory. Many studies are dedicated to searching for compact object in binary systems that contain B-type stars \citep[e.g.][]{2019Natur.575..618L}. To derive precise masses of the compact objects in binary systems, orbital parameters including period, inclination together with mass ratio are essential. Multi-band light curves could help to constrain these parameters. As a result, we plan to carry out multi-band photometric and spectroscopic observations of Be stars.

\subsubsection{Observation design}
Based on previous Be star spectra database \citep[e.g.][]{2011AJ....142..149N} and Be stars identified by the LAMOST sky survey \citep{2022ApJS..260...35W}, we will first search for Be spectra that exhibit variations in radial velocities, which may indicate the existence of secondary companions. Then we will monitor them using the Mini-SiTian telescopes. Similar to the monitoring of flaring stars, Be stars will be observed continuously. In addition, based on the \emph{STRIP} pipeline developed by Gu et al. (2025), if some outbursts appear, immediate spectroscopic observations using the 2.16-m telescope at Xinglong observation will also be applied to monitor the outburst process.

\subsection{Cataclysmic Variable Stars}
\subsubsection{Background}
Cataclysmic variable stars (CVs) represent one kind of objects with very rich variability simultaneously in a wide range of
time scales and amplitudes. They are short orbital period binary systems (typically from 80 minutes to less than 12 hours, but can be as short as several minutes in AM CVn systems) in which a white dwarf (WD) accretes from a low-mass donor that fills its Roche lobe \citep{2003cvs..book.....W}.
The variability of CVs is primarily determined by the mass accretion rate onto the WD and the strength of its magnetic field. Different subtypes of CVs are further classified, including dwarf novae (generating outbursts quasi-periodically) \citep{2020AdSpR..66.1004H}, novalike stars (stay in stable states but occasionally exhibit state transitions) and magnetic CVs (including polars and intermediate polars). 
The mechanisms that trigger the state transitions remain unclear \citep{2020AdSpR..66.1090S}.

Apart from outbursts, CV light curves can display various features such as orbital modulations (eclipses, ellipsoidal modulations, and reflections), superhumps, stellar flares, stochastic flickerings, quasi-period oscillations, and coherent white dwarf spin modulations. Higher-cadence observations are helpful to characterize these variations.
Variability beyond normal dwarf nova outbursts, such as ``micronova'' \citep{2022Natur.604..447S}, magnetically-gated outbursts \citep{2017Natur.552..210S}, ``stunted’’ outbursts \citep{1998AJ....115.2527H}, and anonymous Z Cam outbursts (IW And-type outbursts) \citep{2019PASJ...71...20K,2024MNRAS.531..422S}, have all been discovered in many CVs \citep{2024ApJ...962L..34I}. Multi-band observations are essential for distinguishing the underlying complex physical processes. The orbital period, which is the most fundamental parameter to determine the evolution stage of a CV, requires long-baseline observations for accurate determination. Extending these measurements to a larger sample of CVs will facilitate more comprehensive statistical studies to test CV evolution theories.

\subsubsection{Observation design}
The wide-field Mini-SiTian array and the upcoming SiTian project, which provides simultaneous three-band photometry, will improve the discovery of unidentified cataclysmic variables (CVs). We will analyze light curves to identify new CVs among blue, variable stars, thus enhancing the completeness of known CVs within our local volume \citep{2020MNRAS.494.3799P}. The ongoing Mini-SiTian project has three primary observational goals for studying CVs: (1) Determine the orbital period for more CV candidates, which will utilize the staring mode for CV candidates and poorly known CVs. (2) Capture rapid burst-like events (e.g., micronovae, magnetically gated accretion, stellar flares) with durations of 10 minutes to hours. The staring mode with single exposures under 300 seconds will facilitate detailed studies of the color evolution. (3) Monitor the state transitions among nova-like stars and magnetic CVs. 

The \emph{STRIP} pipeline enables real-time detection of state transitions (e.g., from high to low states) and can trigger follow-up spectroscopic observations (for example, using the Xinglong 2.16-m telescope). These observations aim to explore changes in the accretion disk during such events and determine binary system parameters during phases of low accretion rates. Furthermore, the Mini-SiTian telescopes can also capture possible nova eruptions. The scanning mode is suitable for the discovery of new novae eruptions while the staring mode could monitor known Galactic recurrent novae (e.g., T CrB \citep{2023A&A...680L..18Z}). The Mini-SiTian telescopes together with other telescopes or satellites, such as Einstein-Probe \citep[X-ray;][]{2018SPIE10699E..25Y} and FAST \citep[radio;][]{2011IJMPD..20..989N}, will enable multi-wavelength studies of high-energy phenomena like jets and magnetic accretion in CVs.

\section{Mini-SiTian array, SiTian prototype and beyond}

The Mini-SiTian array has been operated since 2022 and both regular sky surveys and follow-up observations have been conducted. According to data processing results given by Xiao et al. (2025), the Mini-SiTian array is capable of achieving a precision of 5 mmag for stars brighter than 13 mag. Moreover, the \emph{STRIP} pipeline has been used to successfully detect stellar flares on eclipsing binary systems (Gu et al. in 2025). Additionally, the first catalog of variable stars is currently being compiled (Mi et al. in prep). Further intriguing results are anticipated as the Mini-SiTian array continues its operations. We believe that many interesting results will come into being during the future operation of the Mini-SiTian array.

Meanwhile, the SiTian project is advancing steadily. The SiTian prototype, a 1-m telescope, has been installed at the Xinglong Observatory and will be in formal operation earlier 2025. Another SiTian node with three 1-m telescopes at the Lenghu Observatory is currently under construction. With these arrays in place, many new TDEs, SNe, CVs, and stellar flares are expected to be discovered. After the SiTian Project is fully operational, it will produce a huge amount of light curves, which could help us to characterize the variable universe in all aspects.

\acknowledgements

The SiTian project is a next-generation, large-scale time-domain survey designed to build an array of over 60 optical telescopes, primarily located at observatory sites in China. This array will enable single-exposure observations of the entire northern hemisphere night sky with a cadence of only 30-minute, capturing true color (gri) time-series data down to about 21 mag. This project is proposed and led by the National Astronomical Observatories, Chinese Academy of Sciences (NAOC). As the pathfinder for the SiTian project, the Mini-SiTian project utilizes an array of three 30 cm telescopes to simulate a single node of the full SiTian array. The Mini-SiTian has begun its survey since November 2022. The SiTian and Mini-SiTian have been supported from the Strategic Pioneer Program of the Astronomy Large-Scale Scientific Facility, Chinese Academy of Sciences and the Science and Education Integration Funding of University of Chinese Academy of Sciences.
This work is supported by the National Key Basic R$\&$D Program of China via 2023YFA1608303 and the Strategic Priority Research Program of the Chinese Academy of Sciences (XDB0550103).  
This work is supported by the National Natural Science Foundation of China (NSFC; grant Nos. 12090040, 12090041, 12403022, 12422303, 12261141690), the Strategic Priority Research Program of Chinese Academy of Sciences (grant Nos. XDB0550000, XDB0550100 and XDB0550102).

\bibliographystyle{raa}
\bibliography{ms2025-0070}

%
%               one-column-spanning table
%________________________________________ Table 1: Use_of_the routines

\label{lastpage}

\end{document}